# DIC Displacement Measurement Method Based on Improved White Shark Optimizer

## LI Jun[1,2], LEI Zongyu[2],

(1. Hunan Province Engineering Laboratory of Bridge Structure (Changsha University of Science & Technology), Changsha 410114, China; 2. School of Civil Engineering, Changsha University of Science and Technology, Changsha 410114, China)

**Abstract:** The traditional integer-pixel displacement search algorithm of digital image correlation method has low computational efficiency and has been gradually eliminated, and some intelligent optimization algorithms have their own strengths and weaknesses. The white shark optimizer has excellent global search capabilities. However, its calculation is cumbersome, programming complex and inefficient. In order to improve the computational efficiency of the white shark optimizer, it is improved by using the Tent map, introducing the dynamic nonlinear time factor, setting the automatic termination condition and adding the three-step search method. The improved white shark optimizer is applied to the integer-pixel displacement search. In order to improve the accuracy and efficiency of sub-pixel displacement calculation, an improved surface fitting method is proposed by combining bicubic interpolation, improved white shark optimizer and surface fitting method. Through grayscale interpolation, the distance between the fitting points is reduced, and the accuracy of search of the surface fitting method is further improved. The performance of the improved white shark optimizer and the improved surface fitting method is tested by simulated speckle pattern. Test results revealed that the computational efficiency of the improved white shark optimizer is comparable to that of the particle swarm optimization, and the search success rate is as high as 100%. The calculation accuracy of the improved surface fitting method is comparable to that of the Newton-Raphson algorithm, but the computational efficiency is much higher than that of the Newton-Raphson algorithm. Finally, the tensile experiment of low-carbon steel is used to verify the feasibility of the improved surface fitting method in actual measurement.

## Introduction

The Digital Image Correlation (DIC) method is a non-contact optical measurement technique proposed in the 1980s by I. Yamaguchi from Japan, as well as W.H. Peters and W.F. Ranson from the University of South Carolina in the United States[1-2]. By acquiring and analyzing speckle sequence images of the test object before and after deformation, DIC calculates displacement and deformation to obtain physical and mechanical information. It features simple principles, easy experimental implementation, low environmental requirements, high detection accuracy, and full-field real-time detection. With the continuous improvement of camera resolution and DIC theory, DIC has been widely applied in various fields such as civil engineering, machinery, and aerospace[3-6]. Many scholars have made improvements in imaging systems, speckle pattern production, and matching algorithms, enabling DIC to better adapt to different working environments[7,8]. DIC-based displacement measurement mainly includes integer-pixel and sub-pixel displacement measurement, with relatively mature algorithms. Traditional integer-pixel search algorithms include coarse-fine search, hill-climbing search, and cross search; commonly used sub-pixel search algorithms include surface fitting[9],

**Funding:** Supported by the Open Fund of Hunan Provincial Engineering Laboratory for Bridge Structure Safety Control, Changsha University of Science and Technology (Grant No. 14KD07)

**Author Biography:** Li Jun (1972—), female, Master, Senior Experimentalist, currently engaged in research on optical measurement and modern testing technology.
**Email:** lijun@csust.edu.cn

gradient method[10], and Newton-Raphson iteration[11]. With the advancement of camera industrial technology, the number of pixels in images has increased significantly, leading to a proportional increase in the computation time of traditional integer-pixel search algorithms. These algorithms can no longer meet the demand for efficient DIC computation, which restricts the development and application of DIC. The computational accuracy at the integer-pixel level cannot satisfy the requirements of practical engineering, so many scholars focus more on sub-pixel displacement calculation. Among sub-pixel algorithms, the surface fitting method has high computational speed but needs improvement in accuracy; the Newton-Raphson iteration method achieves high accuracy but is time-consuming; the gradient method produces large errors when the actual displacement is far from integer-pixel points. In addition to these classical algorithms, several intelligent optimization algorithms such as genetic algorithm[12], particle swarm optimization (PSO)[13], neural network algorithm[14], and fruit fly optimization algorithm[15] have been introduced into DIC. However, these methods still have room for improvement. For example, PSO is prone to falling into local optima, leading to search failure[16]. The White Shark Optimizer (WSO) is an intelligent optimization algorithm proposed by Braik et al.[17], which exhibits excellent global search capability and is not easily trapped in local optima. Nevertheless, it suffers from complex programming and low computational efficiency. This paper applies an improved WSO algorithm to DIC displacement measurement. Firstly, an integer-pixel displacement calculation method based on the improved WSO algorithm is proposed to quickly and accurately obtain integer-pixel displacement values. Secondly, the existing surface fitting method is improved by combining bicubic interpolation, the improved WSO algorithm, and surface fitting to further reduce the distance between fitting points, thereby enhancing the computational accuracy of the surface fitting method. Numerical simulation results show that the improved algorithm can improve computational efficiency while ensuring accuracy. Tensile test results verify that the improved surface fitting method maintains high accuracy and good stability in practical applications.

## 1 Basic Principles

### 1.1 Digital Image Correlation Method

The working principle of DIC is to capture speckle images of the target object's surface and perform optical tracking of surface displacements. The key to optical tracking is finding the position $(x', y')$ of a point $(x, y)$ from the reference image in the deformed image, as shown in Fig.1. To accurately calculate the displacement of point $(x, y)$, a $(2M + 1) \times (2M + 1)$ region around $(x, y)$ in the reference image is selected as the search window; a $(2N + 1) \times (2N + 1)$ region around $(x, y)$ in the deformed image is used for correlation calculation, where $N > M$. Finally, the extreme point of the correlation coefficient $(x', y')$ is determined, and the distance between $(x', y')$ and $(x, y)$ is the displacement of point $(x, y)$.

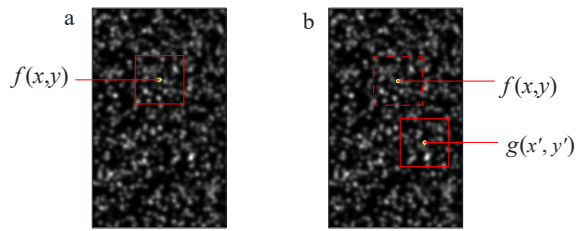

**Fig. 1** Principle of DIC

a—reference image; b—deformed image

To better evaluate the similarity between the reference sub-region and the deformed sub-region, a correlation function should be predefined before correlation matching[18]. The zero-mean normalized

cross-correlation function adopted in this paper is expressed as the following formula (1):

$$C = \frac{\sum\limits_{x=-M}^{M}\sum\limits_{y=-M}^{M}[f(x,y)-f_m][g(x',y')-g_m]}{\sqrt{\sum\limits_{x=-M}^{M}\sum\limits_{y=-M}^{M}[f(x,y)-f_m]^2}\sqrt{\sum\limits_{x=-M}^{M}\sum\limits_{y=-M}^{M}[g(x',y')-g_m]^2}}$$ (1)

where $M$ is the sub-region radius; $f(x, y)$ is the gray value at point $(x, y)$ in the reference image; $g(x', y')$ is the gray value at point $(x', y')$ in the deformed image; $f_m$ and gm are the average gray values of the reference sub-region and the deformed sub-region, respectively.

## 1.2 White Shark Optimizer

As a bionic meta-heuristic algorithm proposed in recent two years, the WSO algorithm has excellent performance in solving complex problems. Based on the behavior of great white sharks in searching for and tracking prey, the WSO algorithm updates the velocity of individual white sharks according to the following formula (2) :

$$v_{k+1}^i = \mu\left[v_k^i + p_1(w_{gbestk}-w_k^i)\times c_1 + p_2(w_{best}^{v_k^i}-w_k^i)\times c_2\right]$$ (2)

where $i = 1, 2, ..., n$ represents the white shark population; $v_k^i$ and $v_{k+1}^i$ are the current and next iteration velocities of individual $i$, respectively; $w_k^i$ is the current position of individual $i$; $w_{gbestk}$ is the best position of individual $i$ in the $k$-th iteration; $w_{best}^{v_k^i}$ is the known best position of the current population; $v^i$ is the index of the best position; $c_1$ and $c_2$ are two random numbers within [0,1]; $p_1$ and $p_2$ are influence parameters; $\mu$ is the contraction factor.

When a great white shark moves towards prey, its new position is defined by the update strategy in the following formula (3):

$$w_{k+1}^i = \begin{cases} w_k^i \cdot \neg \oplus w_0 + u \cdot a + l \cdot b; \ rand < mv \\ w_k^i + v_k^i \ / \ f; \ rand \ge mv \end{cases}$$ (3)

where $w_{k+1}^i$ is the position of individual $i$ in the next iteration; $\neg$ is the negation operator; $\oplus$ is the XOR operation; $w_0$ is a logical vector; $a$ and $b$ are one-dimensional binary vectors; $l$ is the lower limit of the search range; $u$ is the upper limit of the search range; $f$ is the frequency of the individual; rand is a defined random number within [0,1]; $mv$ is the movement force of the individual, which is related to the current number of iterations.

The WSO algorithm uses the following formula (4) to simulate the behavior of white sharks moving towards the white shark closest to the prey:

$$w_{k+1}^{\prime i} = w_{gbestk} + r_1 D_w \, \text{sgn}(r_2 - 0.5) \ ; \ r_3 < S_s$$ (4)

where $w_{k+1}^{\prime i}$ is the new position of individual $i$ in the next iteration; $r_1$, $r_2$, $r_3$ are defined random numbers within [0,1]; sgn is the sign operator; $D_w$ is the distance between the white shark individual and the prey; $S_s$ is the influence parameter of the optimal individual on other individuals.

When other white shark individuals move closer to the optimal white shark individual, their positions are updated according to the following formula (5):

$$w_{k+1}^i = \frac{w_k^i + w_{k+1}^{\prime i}}{2 \times rand} \quad (5)$$

Readers interested in the detailed calculation process of each parameter in the above formulas can refer to Reference [17], which is not repeated here.

## 2 Improved WSO Algorithm

The WSO algorithm has good global search capability and can realize global search in digital image correlation analysis, but there is no application of the WSO algorithm in this field so far. In addition, this method also has shortcomings such as complex calculation process, cumbersome programming, and low calculation efficiency. To improve the computational efficiency of the WSO algorithm, some steps in the algorithm are simplified, retaining only the second terms of Formula (2) and Formula (3). The WSO algorithm is improved by adopting Tent chaos mapping initialization, introducing a dynamic nonlinear time factor, setting iteration termination conditions, and applying the three-step search method.

## 2.1 Tent Chaos Mapping

The Tent chaos mapping can generate random numbers between 0 and 1. Experiments have shown that initializing the population using Tent chaos mapping can make the population distribution more uniform[18]. The mathematical expression of the Tent mapping is as the following formula (6):

$$x_{n+1} = \begin{cases} \dfrac{x_n}{a_0}, & 0 \le x_n \le a_0 \\[2mm] \dfrac{(1-x_n)}{1-a_0}, & a_0 < x_n \le 1 \end{cases} \quad (6)$$

Where $a_0 \in [0,1]$, and $a_0 = 0.5$ is taken. Fig. 2 shows the distribution of the Tent chaos mapping within the range of 0-1.

## 2.2 Dynamic Nonlinear Time Factor

To enable the WSO algorithm to explore more regions in the early stage of iteration and have good local search capability in the later stage of iteration, a dynamic nonlinear time factor $K_t$ is introduced, whose expression is as the following formula (7). The variation curve of $K_t$ is shown in Fig. 3. In the early stage of iteration, a larger $K_t$ value ensures that the white shark population can explore a larger space, avoiding premature convergence to a local optimal solution; in the later stage of iteration, a smaller $K_t$ value can prevent the white shark population from oscillating around the global optimal solution, ensuring the convergence of the algorithm.

$$K_t = \frac{10}{a_1 + e^{a_2(2k/K-1)}} + 0.5 \quad (7)$$

Where $a_1$ and $a_2$ are constants; $a_1 = 6.25$ and $a_2 = 5.0$ are taken; $K$ is the maximum number of iterations.

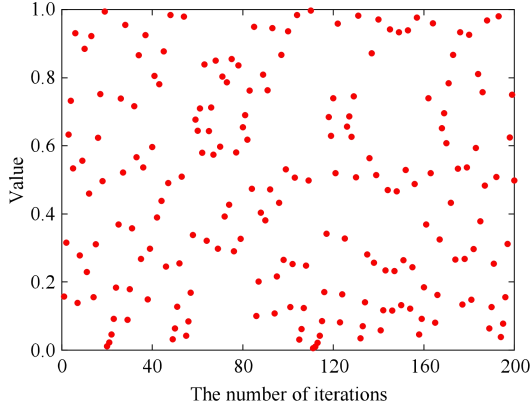
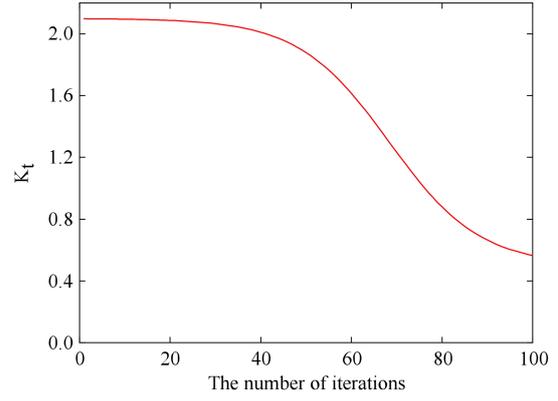

**Fig. 2** Distribution diagram of Tent map    **Fig. 3** The variation curve of dynamic nonlinear time factor Kt

### 2.3 Adaptive Termination Condition

The iteration termination condition given by the WSO algorithm is: terminate the iteration when the number of iterations reaches the threshold. Such a termination condition can better ensure the calculation accuracy when applied in DIC, but if the number of iterations is unreasonably set, meaningless calculations will be repeated, and the efficiency will be seriously reduced. During the search process of the WSO algorithm, the white shark population continuously moves towards the optimal value, and more and more white sharks gather around the global optimal solution. To describe the aggregation degree of white sharks, the population fitness standard deviation (*SD*) is introduced into the WSO algorithm, and its calculation formula is as the following formula (8). The iteration termination condition of the improved WSO algorithm is: terminate the iteration when the population fitness standard deviation reaches the set threshold.

$$SD(k) = \sqrt{\frac{\sum_{i=1}^{n} C(x_i(k) - C_m)^2}{n-1}} \tag{8}$$

Where $C_m$ is the average population fitness; $n$ is the population size; $x_i(k)$ is the coordinate of individual $i$ in the $k$-th iteration. A smaller *SD* value indicates a higher aggregation degree of the population and that individuals are closer to the global optimal point. The threshold is set to 0.001, and the iteration stops when *SD* is less than this value.

### 2.4 Three-Step Search Method

After adopting the above three improvement measures, the global search capability of the WSO algorithm is enhanced, and the search speed is also significantly improved. However, when applied to integer-pixel displacement search in DIC, the global optimal point still cannot be found in individual cases, which will have a great impact on sub-pixel displacement calculation. To ensure the search success rate without affecting the search efficiency, the three-step search method is applied to the WSO algorithm, and its search principle is shown in Fig. 4.

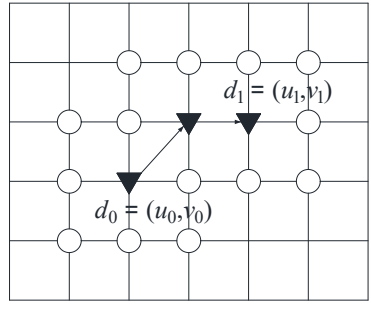

**Fig.4** Principle of three-step search method

The search process is as follows: 1) Taking the solution $d_0$ of the WSO algorithm as the center point, select 8 pixel points closest to $d_0$, and calculate the correlation function of each pixel point. 2) When the position of the maximum value of the correlation function changes, redefine the position of the center point, and calculate the correlation function of the adjacent 8 pixel points again. 3) Repeat step 2) until the position of the maximum value of the correlation function no longer changes.

### 2.5 Integer-Pixel Displacement Calculation Using Improved WSO Algorithm

In the digital image correlation method, the white shark individuals in the improved WSO algorithm are represented as image sub-regions, and the degree of white sharks approaching prey is represented as the correlation coefficient between the reference sub-region and the deformed sub-region. The improved WSO algorithm updates the velocity and position of the sub-region through formula (9) and formula (10).

$$v_i^{k+1} = \mu \left[ v_i^k + p_1(g_{best} - x_i^k) \times c_1 + p_2(p_{i-best} - x_i^k) \times c_2 \right] \tag{9}$$

$$x_i^{k+1} = x_i^k + K_i v_i^{k+1} \tag{10}$$

Where $c_1$ and $c_2$ are two random numbers within [0,1]; $\mu$ is the contraction factor, taking 0.7; $v_i^k$ is the velocity of sub-region $i$ in the $k$-th iteration; $x_i^k$ is the position of sub-region $i$ in the $k$-th iteration; $p_{i-best}$ is the maximum correlation coefficient of the sub-region itself in the $k$-th generation; $g_{best}$ is the global maximum correlation coefficient of all sub-regions in the $k$-th generation; $p_1$ and $p_2$ control the influence of $g_{best}$ and $p_{i-best}$ on $x_i^k$ and the update of sub-region velocity, balancing the global search capability and local search capability of the algorithm, and their calculation methods are shown in formula (11).

$$\begin{cases} p_1 = p_{\max} + (p_{\max} - p_{\min}) \times e^{-(4k/K)^2} \\ p_2 = p_{\min} + (p_{\max} - p_{\min}) \times e^{-(4k/K)^2} \end{cases} \tag{11}$$

Where $p_{\min}$ and $p_{\max}$ represent the initial velocity and subordinate velocity of sub-region movement, taking 0.5 and 1.5 respectively.

The flow chart of integer-pixel displacement search using the improved WSO algorithm is shown in Fig. 5.

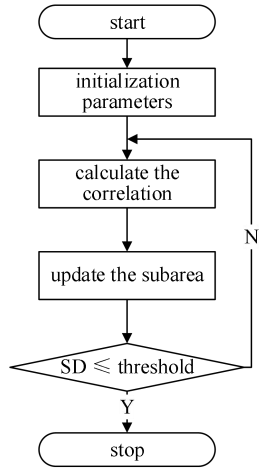

**Fig. 5** Flow chart of improved WSO

The steps for calculating integer-pixel displacement using the improved WSO algorithm are as follows:

1) Initialize parameters: Set the initial parameters of the improved WSO algorithm, including the initial number of sub-regions $N$, the maximum number of iterations $K$, the contraction factor $\mu$, the initial velocity and subordinate velocity $p_{min}$ and $p_{max}$ of sub-region movement, the dynamic nonlinear time factor $K_t$, and the initial position $x$ and velocity $v$ of the sub-region.

2) Calculate correlation coefficient: Calculate the correlation coefficient of each sub-region, and record the global optimal position $g_{best}$ and the local optimal position $p_{i-best}$ of the sub-region.

3) Update sub-region: Update the velocity and position of each sub-region according to formula (9) and formula (10).

4) Terminate iteration: Judge whether the sub-region mean square deviation $SD$ reaches the threshold. If yes, jump out of the loop and output the optimal solution; if no, go to step 2) to continue the iteration. Output the optimal solution when the maximum number of iterations $K$ is reached.

**2.6 Improved Surface Fitting Method**

The improved WSO algorithm proposed above is only used for integer-pixel displacement search, but in practical engineering applications, sub-pixel level accuracy is often required. As one of the classical sub-pixel displacement algorithms, the surface fitting method has the advantages of strong anti-noise ability and fast calculation speed, but its accuracy is lower than that of the gradient method and N-R method. To improve the calculation accuracy of the surface fitting method, a combined method of bicubic interpolation, improved WSO algorithm, and surface fitting is proposed, namely an improved surface fitting method, whose calculation flow is shown in Fig. 6.

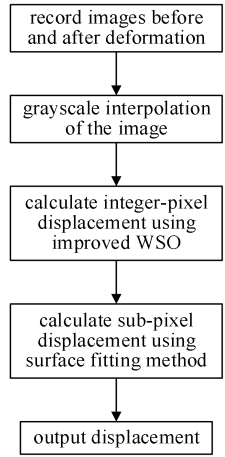

**Fig. 6** Flow chart of improved surface fitting method

The improved surface fitting method inserts 4 points between two integer-pixel points through bicubic interpolation, reduces the distance between fitting points, and improves the integer-pixel displacement search accuracy to 0.2 pixel. After image interpolation, the accuracy of the surface fitting method is easily affected by interpolation errors. Therefore, the calculation errors of different fitting distances are analyzed. Fig. 7 shows the errors when the displacement is 0.2 pixel and 0.8 pixel, and the fitting distance is 0.2~2 pixel.

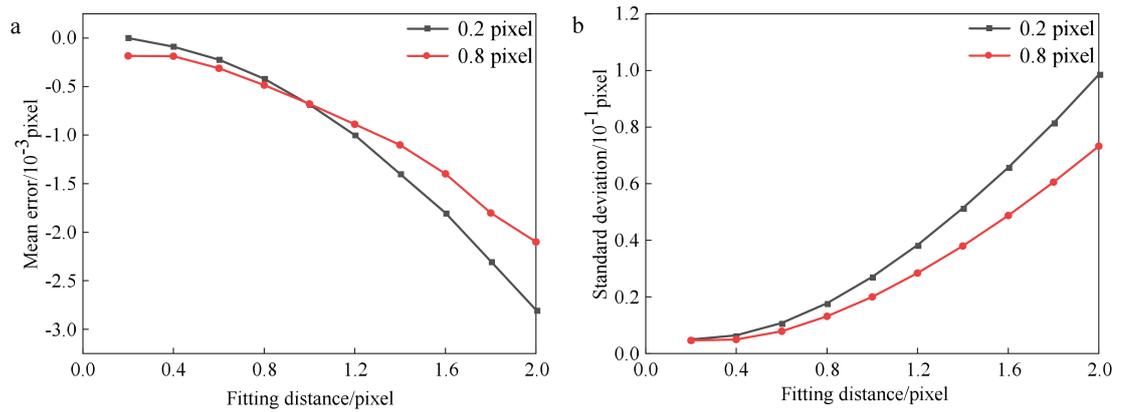

**Fig. 7** Comparison of calculation errors with fitting distances of 0.2~2 pixel

a—mean error; b—standard deviation

It can be seen from Fig. 7 that as the fitting distance increases, the mean error and standard deviation of the improved surface fitting method also increase. Therefore, to reduce calculation errors, the smallest possible fitting distance should be selected. The fitting distance adopted in this paper is 0.2 pixel.

## 3 Numerical Simulation

In this study, MATLAB software was used to generate simulated speckle patterns following the method proposed by Peng Zhou et al.[20]. The image size is 256 pixel × 256 pixel, with 1000 speckles each of 4 pixel in size. The speckle images before and after deformation are shown in Fig. 8.

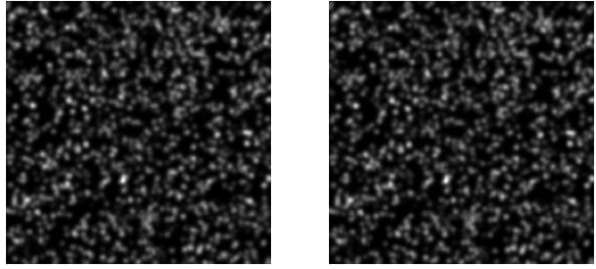

**Fig. 8** Speckle pattern

a—before deformation; b—after deformation

### 3.1 Integer-Pixel Search Success Rate and Computational Time

To verify the performance of the improved WSO algorithm in integer-pixel displacement search, reference images were generated using the above method, and a series of deformed images with displacements ranging from 1 to 10 pixel were obtained through translation operations. The coarse-fine search method, PSO algorithm, original WSO algorithm, and improved WSO algorithm were respectively applied to calculate integer-pixel displacements. Comparisons of their search success rates and computational times are presented in Fig. 9 and Fig. 10.

As indicated in Fig. 9 and Fig. 10, the coarse-fine search method and original WSO algorithm achieve a search success rate close to 100% but exhibit low computational efficiency. Although the PSO algorithm offers fast computation speed, its search success rate is only around 80%. The improved WSO algorithm maintains a computational success comparable to the PSO algorithm while achieving a search success rate as high as 100%. This demonstrates that the improved WSO algorithm has significantly enhanced computational efficiency compared to the original WSO algorithm. Meanwhile, it improves the global search capability for digital images and avoids falling into local optima, a common issue with the PSO algorithm, thereby providing accurate and reliable initial values for subsequent sub-pixel displacement search.

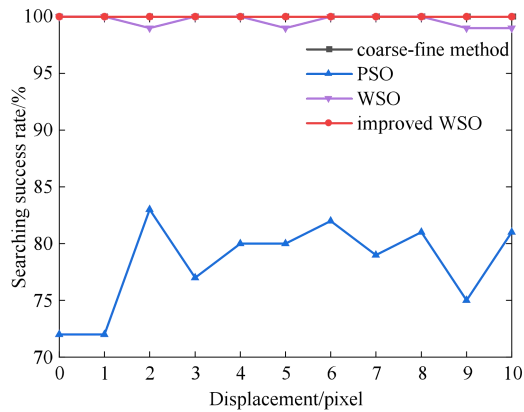

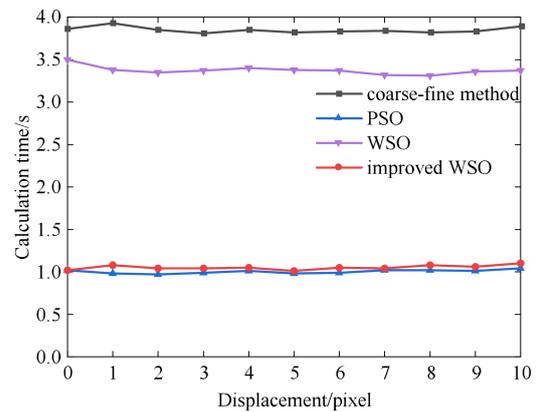

**Fig. 9** Search success rate of four algorithms    **Fig. 10** Computational time of four algorithms

### 3.2 Sub-Pixel Displacement Search Error Analysis

To accurately evaluate the computational accuracy of the improved surface fitting method, the reference image was translated to generate a series of deformed images with sub-pixel-level micro-deformations (displacements ranging from 0.05 to 1 pixel). The improved surface fitting method, original surface fitting method, gradient method, and Newton-Raphson (N-R) method were used to calculate the displacements of the target image set. The mean errors and standard deviations of the four algorithms are shown in Fig. 11.

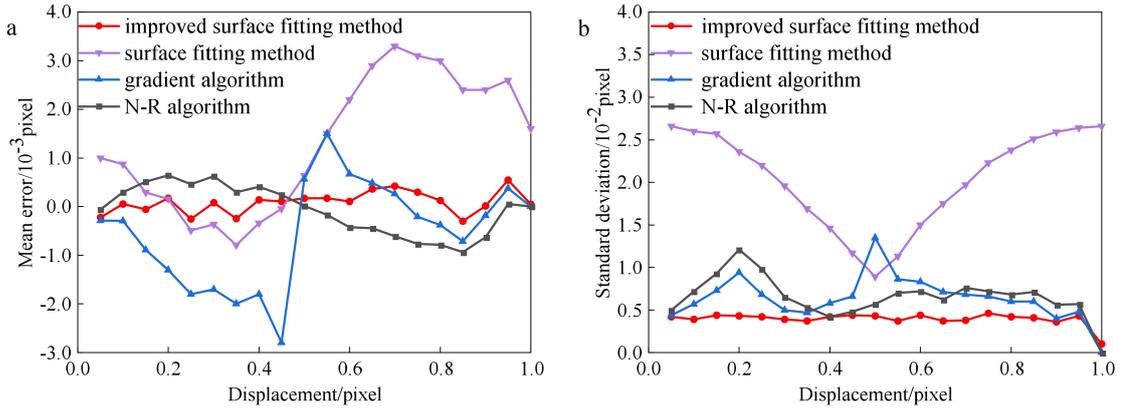

**Fig. 11** Comparison of errors among four algorithms

a—mean error; b—standard deviation

As shown in Fig. 11, the N-R method exhibits small mean error and standard deviation with stable fluctuations. The gradient method shows significant fluctuations around 0.5 pixel, which is attributed to the large error caused by neglecting high-order infinitesimals in the first-order Taylor series expansion[21]. The original surface fitting method has large fluctuations in both mean error and standard deviation, but achieves a small mean error and minimum standard deviation at a displacement of 0.5 pixel. The improved surface fitting method has smaller mean error and standard deviation than the other three algorithms, indicating lower error and excellent stability.

To verify the computational efficiency of the improved surface fitting method, Table 1 summarizes the computational times of the N-R method, gradient method, original surface fitting method, and improved surface fitting method.

**Table 1** Computational time of four algorithms

| algorithm | N-R algorithm | gradient algorithm | surface fitting method | improved surface fitting method |
|---|---|---|---|---|
| time/s | 55.64 | 3.84 | 3.31 | 1.95 |

As shown in Table 1, the N-R method, despite its high sub-pixel accuracy, has much lower computational efficiency than the other methods. The original surface fitting method consumes slightly less time than the gradient method, but the difference is not significant. Although the improved surface fitting method involves bicubic interpolation, it achieves much higher computational efficiency than the other three algorithms due to the fast computation speed of the improved WSO algorithm.

To investigate the smoothness of displacement fields generated by sub-pixel algorithms, the reference image was translated by 0.5 pixel, and displacement fields were generated using the four algorithms mentioned above. The results are shown in Fig. 12.

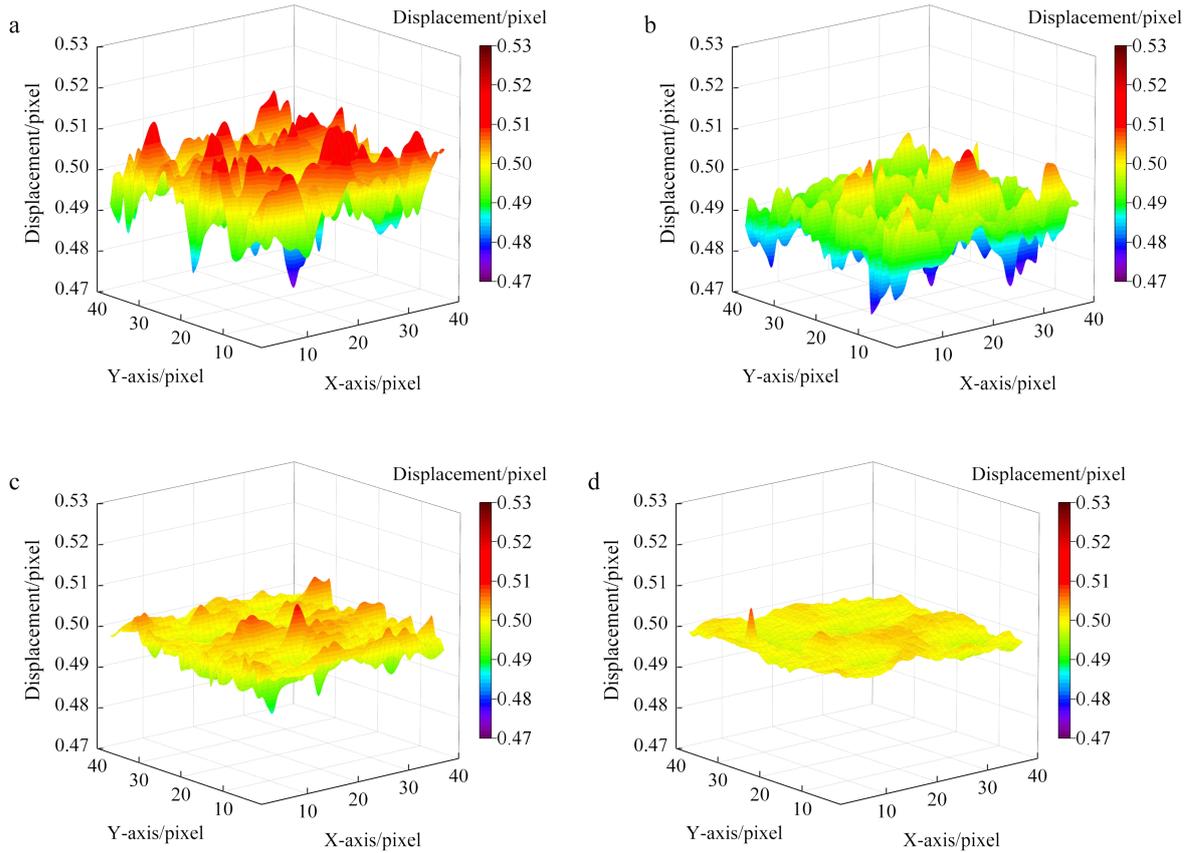

**Fig. 12** Displacement fields generated by four algorithms

a—N-R algorithm; b—gradient algorithm; c—surface fitting method; d—improved surface fitting method

As observed in Fig. 12, the displacement fields generated by the gradient method and N-R method have large oscillation amplitudes and poor smoothness. The original surface fitting method produces a relatively stable displacement field since its standard deviation reaches the minimum at 0.5 pixel. The displacement field generated by the improved surface fitting method is the smoothest, with errors controlled within an ideal range, demonstrating its superior performance in displacement field calculation.

## 4 Experimental Verification

Numerical simulations only analyze algorithm performance under theoretical conditions without considering the influence of external environmental factors such as noise. To verify the effectiveness of the improved surface fitting method in practical measurements, a tensile test was conducted on a low-carbon steel specimen (250 mm in length, 30 mm in width, 5 mm in thickness) with surface-sprayed speckles. A CCD camera and a binocular integrated camera were used to capture speckle images during the tensile process. The improved surface fitting method and the other three classical sub-pixel algorithms were applied to calculate the displacements of the deformed images, with comparisons made against the commercial DIC software VIC-3D. In this experiment, the $v$-displacement field of the Region of Interest (ROI) of the specimen (shown in Fig. 13) and the $y$-direction displacement of point $P_0$ were measured.

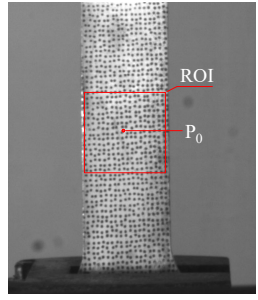

**Fig.** 13 Tensile specimen and the ROI area

To study the performance of sub-pixel algorithms in practical measurements, the *v*-displacement fields of the ROI were calculated using the four algorithms, and the results are shown in Fig. 14.

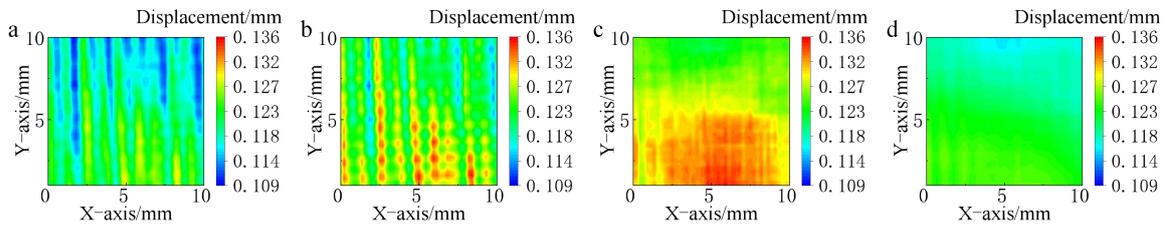

Fig. 14 v-displacement fields calculated by four algorithms

a—N-R algorithm; b—gradient algorithm; c—surface fitting method; d—improved surface fitting method

As shown in Fig. 14, the three classical algorithms exhibit poor stability with large fluctuations in their displacement fields. Specifically, the displacement fields generated by the N-R method and gradient method show obvious local fluctuations, while the original surface fitting method produces displacement fields with significant jumps. In contrast, the displacement field generated by the improved surface fitting method is smooth and uniformly distributed without local fluctuations. To evaluate the accuracy of sub-pixel algorithms in practical measurements, 10 images were selected from the series of deformed images. The *y*-direction displacement *v* of point $P_0$ was calculated using the four sub-pixel algorithms and VIC-3D software. The relative error between each algorithm's results and the VIC-3D measurements was computed, and the average of the 10 relative errors is presented in Table 2.

**Table 2** Relative errors between the four algorithms and VIC-3D

| algorithm | N-R algorithm | gradient algorithm | surface fitting method | improved surface fitting method |
|---|---|---|---|---|
| relative error average/% | 4.04 | 4.96 | 6.06 | 4.13 |

As shown in Table 2, the relative errors between the four sub-pixel algorithms and VIC-3D range from 4.04% to 6.06%. The original surface fitting method has the largest relative error, followed by the gradient method with a slightly smaller error. The improved surface fitting method achieves an accuracy comparable to the N-R method. Consistent with the numerical simulation results, these experimental findings confirm that the improved surface fitting method maintains excellent performance in practical measurements.

## 5 Conclusions

A digital image correlation displacement measurement method based on an improved White Shark Optimizer (WSO) is proposed. To address the low computational efficiency of the original WSO algorithm when applied to DIC displacement calculation, several improvements are implemented: Tent chaos mapping initialization is adopted to achieve a more uniform initial population distribution；A dynamic nonlinear time factor is introduced to balance the exploration and exploitation capabilities of the algorithm； An adaptive iteration termination condition is set to avoid unnecessary redundant computations；The three-step search method is integrated to further improve the algorithm's accuracy. Numerical simulation results demonstrate that compared with the traditional coarse-fine search method, PSO algorithm, and original WSO algorithm, the improved WSO algorithm enhances integer-pixel displacement calculation capability, achieving higher search success rate and computational efficiency.

To improve sub-pixel displacement calculation accuracy, an improved surface fitting method is proposed based on the improved WSO algorithm. Numerical simulations using simulated speckle patterns and comparative analysis with common algorithms show that the improved surface fitting method achieves accuracy comparable to the N-R method while requiring only 3.50% of the computational time of the N-R method. Additionally, tensile tests on low-carbon steel specimens verify the excellent performance of the improved surface fitting method in practical applications. In summary, the improved surface fitting method exhibits outstanding performance in both computational accuracy and efficiency, meeting the requirements of practical engineering. The introduction and improvement of existing algorithms in this study provide an effective alternative for DIC displacement algorithms, which is of great significance for the research and application of digital image correlation methods.

## References

[1] YAMAGUCHI I. A laser-speckle strain gauge[J]. Journal of Physics E: Scientific Instruments, 1981, 14(11): 1270-1273.

[2] PETERS W H, RANSON W F. Digital imaging techniques in experimental stress analysis[J]. Optical Engineering, 1982, 21(3): 427-431.

[3] PETERS W H, RANSON W F, SUTTON M A, et al. Application of digital correlation methods to rigid body mechanics[J]. Optical Engineering, 1983, 22(6): 738-742.

[4] WEI B, LIANG J, LI J, et al. 3D full-field wing deformation measurement method for large high-wing airplanes[J]. Acta Aeronautica et Astronautica Sinica, 2017, 38(7): 172-181 (in Chinese).

[5] GE P X, LI G H. Application of a combinatorial DIC algorithm in sub-pixel displacement measurement[J]. Laser & Optoelectronics Progress, 2018, 55(11): 111202 (in Chinese).

[6] [6] WU R, LIU Y, ZHOU J M. Full-field deformation measurement of wind turbine blades using digital image correlation[J]. Chinese Journal of Scientific Instrument, 2018, 39(11): 258-264 (in Chinese).

[7] WANG J S, ZHANG W J, PAN Z W, et al. Thermal strain measurement based on laser speckle digital image correlation method[J]. Laser Technology, 2023, 47(2): 171-177 (in Chinese).

[8] YE J C, JI H W. A digital image correlation method based on pre-deformation forward additive Gauss-Newton method[J]. Journal of Experimental Mechanics, 2022, 37(5): 621-628 (in Chinese).

[9] LI K Q, ZHU D, DONG X X. Digital speckle correlation method based on improved curved surface fitting method[J]. Laser & Optoelectronics Progress, 2018, 55(5): 051001 (in Chinese).

[10] LI K P, CAI P. Study on the performance of sub-pixel algorithm for digital image correlation[J]. Chinese Journal of Scientific Instrument, 2020, 41(8): 180-187 (in Chinese).

[11] AKBAS S D. Nonlinear thermal displacements of laminated composite beams[J]. Coupled Systems Mechanics, 2018, 7(6): 691-705.

[12] LIU Y, XIAO S D, ZHANG R, et al. Initial estimation of digital image correlated deformation based on genetic algorithms[J]. Laser Technology, 2020, 44(1): 130-135 (in Chinese).

[13] WANG Y H, ZHANG H, CHEN L, et al. Adaptive decision inertia weight PSO correlation searching algorithm[J]. Opto-Electronic Engineering, 2015, 42(8): 1-7 (in Chinese).

[14] SUN H B. Study on identification recognition algorithm based on artificial neural network[J]. Electronic Technology & Software Engineering, 2021(10): 115-116 (in Chinese).

[15] ZHANG C H. Related technical research on integer-pixel displacement measurement of digital speckle image[D]. Taiyuan: North University of China, 2016 (in Chinese).

[16] CHEW K S, ZARRABI K. Non-contact displacements measurement using an improved particle swarm optimization based digital speckle correlation method[C]//Proceedings of the 2011 International Conference on Pattern Analysis and Intelligence Robotics. Kuala Lumpur, Malaysia: IEEE, 2011: 53-58.

[17] BRAIK M, HAMMOURI A, ATWAN J, et al. White Shark Optimizer: A novel bio-inspired meta-heuristic algorithm for global optimization problems[J]. Knowledge-Based Systems, 2022, 243: 108457.

[18] MA S P, JIN G C. New correlation coefficients designed for digital speckle correlation method (DSCM)[J]. Proceedings of SPIE, 2003, 5058: 25-33.

[19] HUANG M L, ZHAO Z J, PU L N, et al. Particle swarm optimization algorithm based on adaptive Tent chaos search[J]. Journal of Computer Applications, 2011, 31(2): 485-489 (in Chinese).

[20] ZHOU P. Subpixel displacement and deformation gradient measurement using digital image/speckle correlation (DISC)[J]. Optical Engineering, 2001, 40(8): 1613-1620.

[21] DONG Y. Research on sub-pixel plane displacement measurement based on digital speckle correlation method[D]. Hangzhou: China Jiliang University, 2021 (in Chinese).

# DIC Displacement Measurement Method Based on Improved White Shark Optimizer

*LI Jun[1,2], LEI Zongyu[2],*

(1. Hunan Province Engineering Laboratory of Bridge Structure (Changsha University of Science & Technology), Changsha 410114, China; 2. School of Civil Engineering, Changsha University of Science and Technology, Changsha 410114, China)

**Abstract:** Nowadays, digital image correlation method has become a key focus in many industries. With the continuous improvement of the basic theory of DIC and the rapid development of camera technology, it is widely used in civil engineering, mechanical engineering, aerospace and other fields. The DIC method provides many scholars with a convenient, low-cost, and high-precision method for measuring displacement and strain. With the development of DIC, people have higher and higher requirements for it, and how to improve its accuracy and efficiency is a problem that many scholars have been thinking about.

As a non-contact optical measurement method, DIC is welcomed by many researchers due to its high accuracy, simple operation, and full-field measurement. In general, displacement measurement using DIC is mainly divided into two steps: integer-pixel displacement search and sub-pixel displacement search. The integer-pixel displacement search is the premise of sub-pixel displacement search, and the sub-pixel displacement search can better meet the needs of practical engineering. The traditional integer-pixel displacement measurement methods include: coarse-fine search method, hill climbing search method, cross search method, etc., among which the coarse-fine search method has the lowest computational efficiency, and the other two algorithms have slightly higher computational efficiency than the coarse-fine search method, however, these traditional integer-pixel displacement measurement methods have gradually failed to meet the requirements of researchers. Therefore, some intelligent optimization algorithms are applied to the measurement of integer-pixel displacement, such as particle swarm optimization algorithm, genetic algorithm, neural network algorithm, etc. The introduction of some intelligent optimization algorithms has greatly improved the speed of integer-pixel displacement search, but these algorithms also have some drawbacks. For example, genetic algorithms are complex to calculate and difficult to program and the particle swarm optimization is susceptible to the influence of local optimums and neural networks require large amounts of data to train. The white shark optimizer has a strong global search ability, which is not easy to be affected by the local optimal value, and has good performance in solving extreme values. Therefore, this paper introduces the white shark optimizer into the integer-pixel displacement search of DIC, but it is found that the computational efficiency of the white shark optimizer is slower, which is only slightly better than the coarse-fine search method. In order to improve the computational efficiency of the white shark optimizer, the following aspects are improved: the introduction of Tent map, the introduction of dynamic nonlinear time factor, the setting of automatic termination conditions and the addition of a three-step search method. The improved white shark optimizer not only improves the computational efficiency, but also can effectively ensure the accuracy of the integer-pixel displacement search. It and can realize the accurate measurement of the integer-pixel displacement.

The accuracy of displacement measurement at the integer-pixel level is often not satisfactory to researchers, so we need to achieve displacement measurement at the sub-pixel level on the basis of integer-pixel displacement. The traditional sub-pixel displacement measurement methods include surface fitting method, gradient method, and Newton-Raphson. The surface fitting method is the most efficient method, followed by the gradient method, and the Newton-Raphson consumes a lot of time. The calculation accuracy of the Newton-Raphson is much higher than that of the surface fitting method and the gradient method. These traditional sub-pixel displacement search algorithms usually cannot take into account both accuracy and efficiency, and we need to rely on our own needs to select the appropriate algorithm. In order to further improve the accuracy and efficiency of sub-pixel search, this paper combines the bicubic interpolation, the improved white shark optimizer and the surface fitting method, and proposes an improved surface fitting method. The improved fitting method reduces the distance between the fitting points through grayscale interpolation, and further improves the search accuracy of the surface fitting method. In addition, the white shark optimization algorithm is used to search for integer-pixel displacement, which ensures that the algorithm has high efficiency.

The performance of the proposed algorithm is tested by simulating speckle pattern, The mean error and standard deviation of the surface fitting method, the gradient method, the Newton-Raphson, and the improved surface fitting method is tested when the displacement is 0-1 pixel. In addition, when the displacement is 0.5 pixels, the displacement field of the four algorithms is displayed. The results reveal that the improved surface fitting method not only has a much higher calculation accuracy than the surface fitting method, but also is comparable to the Newton-Raphson. In terms of computational speed, the computational efficiency of the improved surface fitting method is not only much higher than that of the Newton-Raphson iterative method, but also slightly higher than that of the surface fitting method. In this paper, the improved white shark optimizer is introduced into DIC, and the integer-pixel search and sub-pixel search are improved. Finally, we obtained a high-efficiency and high-precision DIC displacement search algorithm, which made some contributions to the development of DIC.

**Key words:** measurement and metrology; digital image correlation method; white shark optimizer; surface fitting method; displacement measurement

------------------------

**Biographies**

Li Jun (1972 - ), female, Han nationality, Master of Engineering, Senior Experimentalist. Her main research interests focus on optical measurement and modern testing technology.

Lei Zongyu (2001 - ), male, Zhuang nationality, Master's degree holder. His main research interests focus on digital image correlation and optical measurement.

**Affiliations**

1. Hunan Provincial Engineering Laboratory for Bridge Structure Safety Control (Changsha University of Science and Technology);

2. Changsha University of Science and Technology

**Address**

Changsha University of Science and Technology, No. 960, Section 2 of South Wanjiuli Road, Muyun Subdistrict, Tianxin District, Changsha City, Hunan Province, China

Tel: 13319588653

E-mail: lijun@csust.edu.cn

**Innovation Points**

(1) It is found through research that the White Shark Optimizer (WSO) exhibits extremely high accuracy but low computational efficiency. This study adopts multiple measures to improve the WSO algorithm, significantly enhancing its computational efficiency while maintaining high accuracy. (2) The improved WSO algorithm is applied to integer-pixel displacement search in Digital Image Correlation (DIC), enabling fast and accurate integer-pixel displacement retrieval. Numerical simulation results demonstrate that the performance of the improved WSO algorithm is superior to that of the traditional coarse-fine search method and Particle Swarm Optimization (PSO) algorithm. (3) To enhance the performance of the surface fitting method in sub-pixel displacement calculation, a novel improved surface fitting method is proposed by combining bicubic interpolation, the improved WSO algorithm, and the traditional surface fitting method. Both numerical simulations and low-carbon steel tensile tests verify that the improved surface fitting method achieves both high computational accuracy and efficiency.